\documentclass[aps, prb, twocolumn, showkeys]{revtex4-1}

\usepackage{chemformula} % for chemical symbols and formulae
\usepackage{siunitx} % for si units
\usepackage[inline,shortlabels]{enumitem} % inline numbered lists
\usepackage{hyperref} %hyperref can cause crash if citation is split over two pages. This is remedied by the [draft] option, which can be removed after rearranging the text a bit.
\usepackage{graphicx} % including graphics
\usepackage{natbib}
\begin{document}

    \title{Interplay between magnetism and Na concentration in \ch{Na_{x}CoO2}}
    \author{M. H. N. Assadi}
    \email{assadi@aquarius.mp.es.osaka-u.ac.jp}
    \author{H. Katayama-Yoshida}
    \affiliation{Graduate School of Engineering Science, Osaka University, Osaka 560-8531, Japan}

    \begin{abstract}
        Through comprehensive density functional calculations, the crystallographic, magnetic and electronic properties of \ch{Na_{x}CoO2} ($x = 1$, 0.875, 0.75, 0.625 and 0.50) were investigated. We found that all Na ions in \ch{Na_{}CoO2} and \ch{Na_{0.875}CoO2} share the basal coordinates with O ions. However, as $x$ decreases, some of Na ions move within the basal plane in order to reduce the in-plane \ch{Na}--\ch{Na} electrostatic repulsion. Magnetically, there was strong tendency for type A antiferromagnetism in the \ch{Na_{0.75}CoO2} system, while all other Na deficient systems had a weaker ferromagnetic tendency. The results on magnetism were in excellent agreement with the experiments.
    \end{abstract}
    \keywords{Sodium Cobaltate; Density Functional Theory; Magnetism.}
    \date{2015}
    \maketitle

    Functional \ch{Na_{x}CoO2} has a diverse range of magnetic, superconducting and electronic phases that can be tuned by varying the Na concentration ($x$).\cite{Koumoto2006}
    For instance, \ch{Na_{x}CoO2} has been shown to be a promising material for high efficiency thermoelectric systems.\cite{Fergus2012}
    The thermoelectric performance of \ch{Na_{x}CoO2} is caused by high degree of electronic frustration that results in large spin entropy and thus large Seebeck coefficient.\cite{Terasaki1997, Wang2003}
    Additionally, in \ch{Na_{x}CoO2}, phonons are strongly scattered by the \ch{Na+} ions which are, to a great extent, fragmentarily ordered at room temperature.
    This combination leads to unprecedented freedom to favorably adjust all interdependent factors of the thermoelectric figure of merit ($ZT$) more independently when compared to other thermoelectric materials.
    Furthermore, as an ionic oxide \ch{Na_{x}CoO2}, has advantage for operations at higher temperatures and oxidizing environment.\cite{Assadi2013}

    From materials engineering viewpoint, varying \ch{Na_{x}CoO2}'s Na concentration has been the primary technique to push the $ZT$ to higher limits or to manipulate other functional properties.
    \ch{Na_{x}CoO2}'s layered lattice is made of alternating Na and edge-sharing \ch{CoO6} octahedral layers.
    Na ions can occupy either Na1 or Na2 sites which each share their basal coordinates with Co and O respectively.
    Intuitively, Na occupation of Na1 site seems unfavorable due to high electrostatic repulsion between Co and Na ions.
    However, Na1 site occupation may become favorable in lower Na concentrations when Na ions can decrease the in-plane electrostatic energy by moving to Na1 site and therefore increasing the \ch{Na}--\ch{Na} distance.
    In the past few years there have been several experimental\cite{Morris2007, Roger2007} and computational\cite{Zhang2005, Meng2008, Wang2007} attempts to establish the exact Na ordering in the \ch{Na_{x}CoO2}.
    However, a unifying picture that relates the Na ordering to the magnetic interaction in \ch{Na_{x}CoO2} is yet to be presented.
    In this work, we investigate the Na ordering patterns and with emphasis on their implications on the magnetic behaviour of the \ch{Na_{x}CoO2} systems for Na concentrations of $x = 0.50$, 0.625, 0.75, 0.875 and 1.00 by density functional theory (DFT).
    \ch{Na_{x}CoO2} with higher Na content of $x > 0.50$, as investigated here, possesses excessively higher thermopower thus it is an appealing for energy generation applications.

    We first optimized the lattice parameters of the \ch{Na_{}CoO2} primitive cell and obtained the theoretical lattice parameters of $2.87\si{.\angstrom}$ for $a$ and $10.90\si{.\angstrom}$ for $c$ which were in reasonable agreement with the single crystal lattice parameters\cite{Voneshen2013} differing by $1.04\%$ and $0.91\%$ for $a$ and $c$ respectively.
    The computational settings are presented in details in the Supplementary Information.
    We then constructed a $4a\times2a\times1c$ supercell of \ch{Na_{}CoO2} consisting of 64 ions to study the Na deficient systems.
    Na vacancies were created by subsequently removing Na ion from the structure to reach the desired Na concentrations.
    There were numerous possible arrangements for Na ions in Na deficient system, many related to each other by symmetry considerations.
    Therefore, in search for the ground state configuration, for any given Na concentration we calculated the total energy of various Na patterns that were specifically created by varying the ratio of Na1/Na2 Na ions.

    \begin{figure}
        \centering
        \includegraphics[width=0.95\columnwidth]{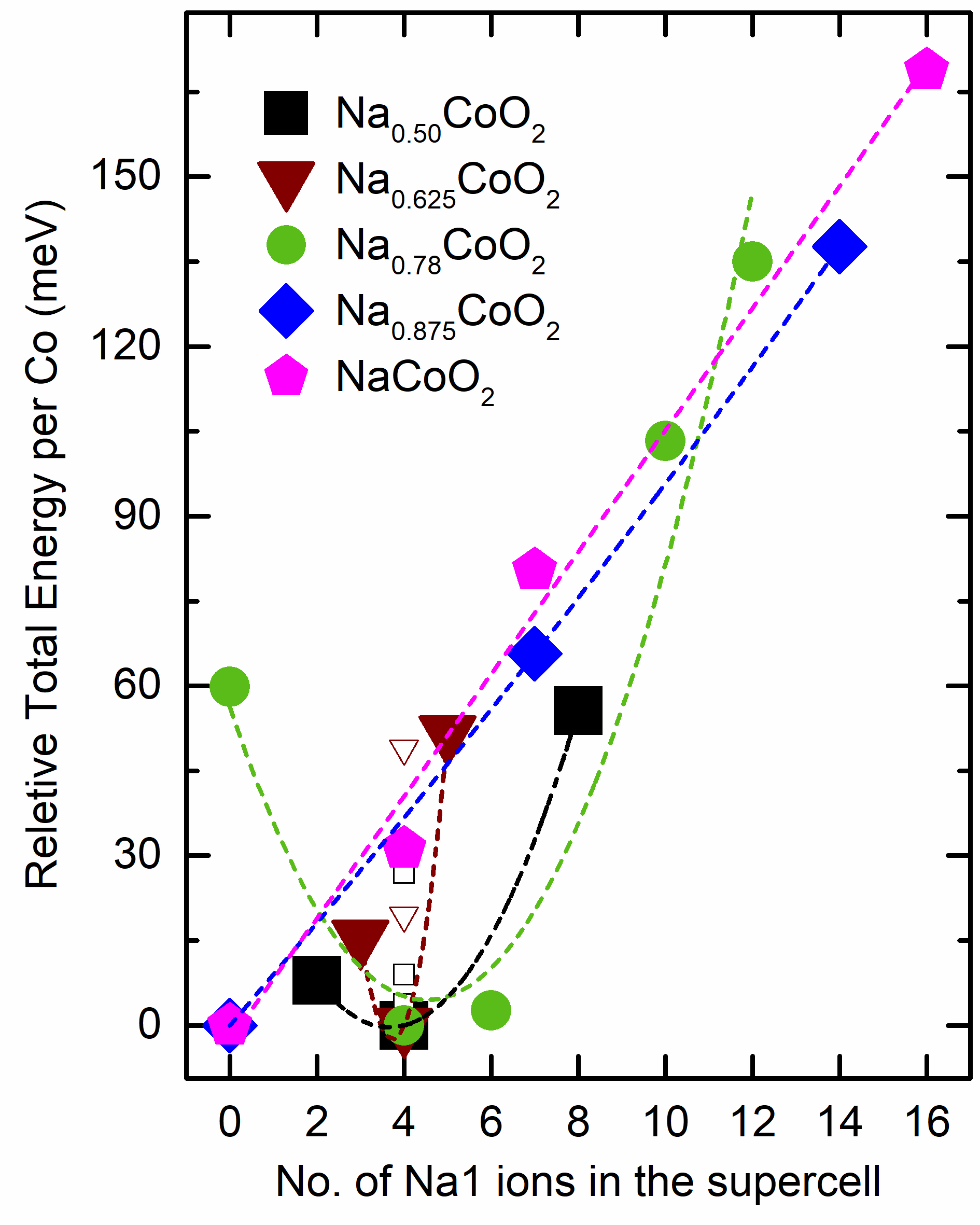}
        \caption{\label{fig:1}The total energy of different Na ordering pattern in \ch{Na_{x}CoO2} systems. The energies are presented relative to the lowest energy pattern for any given Na concentration. The \ch{Na_{}CoO2}, \ch{Na_{8.875}CoO2}, \ch{Na_{0.75}CoO2}, \ch{Na_{0.625}CoO2} and \ch{Na_{0.5}CoO2} systems are represented by pink, blue, green, brown and black symbols respectively. The empty symbols represent those pattern of higher total energy for a given number of Na1 ions.}
    \end{figure}

    To obtain the final Na pattern, we allowed the internal coordinates of all ions in the supercell to relax while fixing the lattice constants to the theoretical values of the pristine \ch{Na_{}CoO2}.
    This procedure guarantees that artificial hydrostatic pressure is avoided in finding the final optimized configurations.
    The total energy of the calculated configurations versus the number of Na1 ions in the supercell is presented in Fig. \ref{fig:1}.
    The dashed lines were drawn to illustrate the relative stability of the ground state configurations with respect to configurations with different number of Na ions at Na1 coordinates.
    Furthermore, we normalized the presented total energies per number of primitive cells per supercell.
    This is to allow objective comparison of the trends across systems with different Na concentration.

\begin{figure}
        \centering
        \includegraphics[width=0.95\columnwidth]{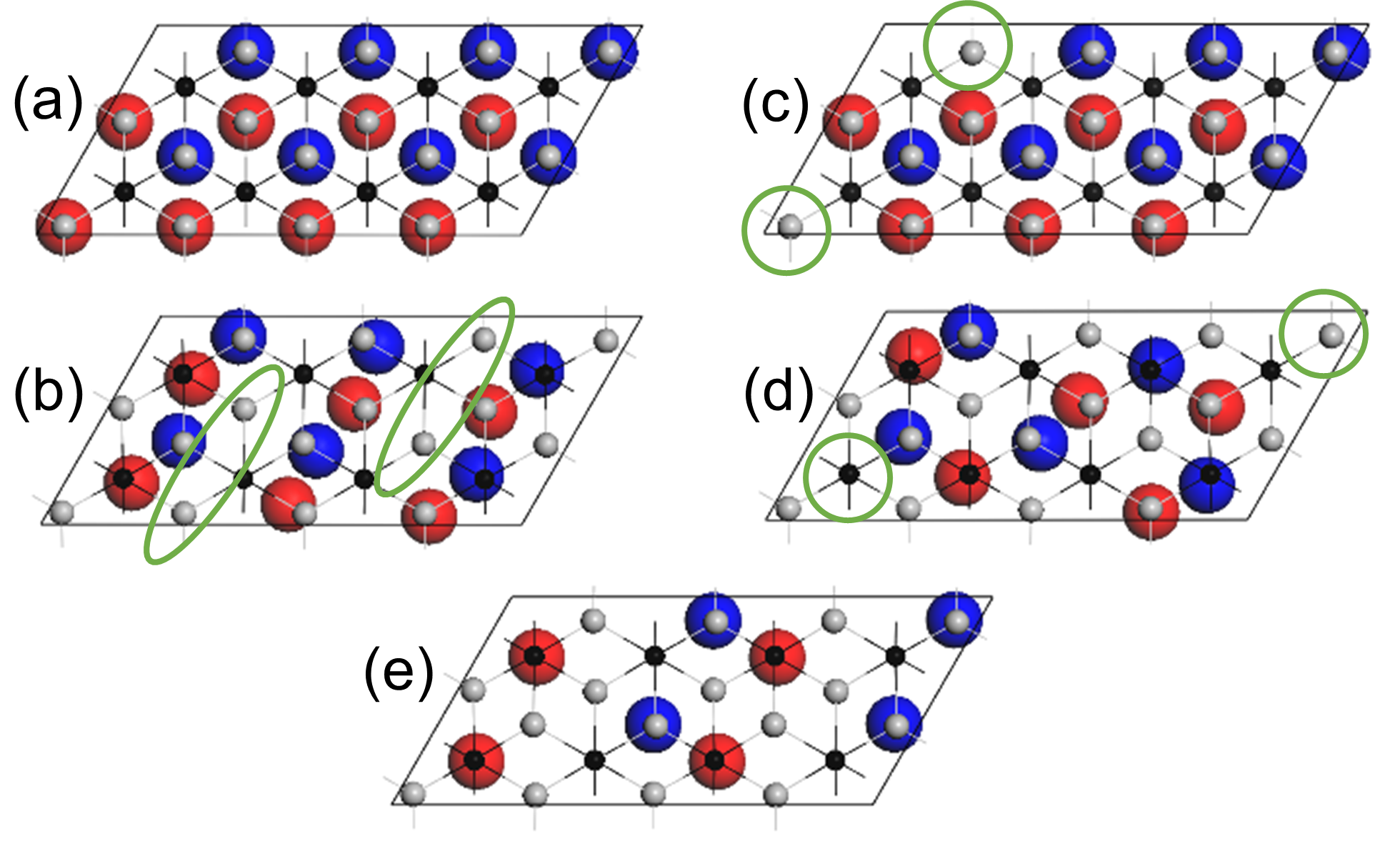}
        \caption{\label{fig:2}The positions of the Na+ ions in the supercell of the (a) \ch{Na_{}CoO2}, (b) \ch{Na_{5}CoO2} and (c) \ch{Na_{5}CoO2}, (d) \ch{Na_{5}CoO2} and (e) \ch{Na_{0}CoO2} systems. The rigid black and gray spheres represent Co and O ions respectively. The larger red and blue spheres represent \ch{Na+} ions with fractional coordinate of Z = 0 and Z = 0.5 respectively.}
    \end{figure}

    We first examined the Na ordering pattern in \ch{Na_{}CoO2} with full Na occupancy.
    In this case, when all Na ions occupy either Na1 or Na2 sites, the structure has the highest possible symmetry, which may hint for the most stable Na ordering.
    We found that the latter case (all Na ions on Na2 sites) was the most stable configuration having a lower total energy of $168\si{.meV/Co}$ than that of a configuration in which all Na ions were placed at Na1 sites.
    A schematic representation of this structure is shown in Fig. \ref{fig:2}(a) in which all Na ions share their basal Cartesian coordinates with O ions.
    Here the shortest distance between the Na ions in the optimized structure was $2.87\si{.\angstrom}$ and which is equal to the lattice parameter a indicating the full preservation of symmetry of the primitive cell which is hexagonal of P6(3) space group.
    This Na ordering pattern is in agreement with earlier DFT calculations and experimental observations.\cite{Zhang2005, Meng2008}
    In the next stage, the Na ions' pattern in Na deficient \ch{Na_{x}CoO2} systems was studied.
    For the \ch{Na_{0.875}CoO2} system, two Na vacancies were created by removing one Na ion from the top Na layer (Z=0.5) and another Na ion from the bottom Na layer (Z=0) resulting in $87.5\%$ Na concentration.
    As shown in Fig. \ref{fig:2}(b), in the \ch{Na_{0.875}CoO2} system, all Na ions were found to occupy Na2 sites.
    Since there was one Na vacancy in each layer, Na ions relaxed to a less compact pattern to reduce the electrostatic repulsion in the Na layers. However, the shortest distance between Na ions remained $2.87\si{.\angstrom}$ as the same as in \ch{Na_{}CoO2} system.
    The introduction of the vacancies impaired the high symmetry of the pristine \ch{Na_{}CoO2} structure.
    Therefore, the relaxed \ch{Na_{0.875}CoO2} structure was instead orthorhombic of PMMA space group --eight symmetry operations--.
    One notable feature of this structure was the tendency of Na ions to occupy Na2 sites.
    We found that the energy penalty for moving Na ions from Na2 sites to Na1 sites was rather high.
    For example, for a configuration in which half of the Na ions on Na1 sites, the total energy was higher by $\sim 65 \si{.meV/Co}$, while if all Na ions were to be moved to Na1 sites the total energy would be higher by $\sim 137 \si{.meV/Co}$.
    Previous DFT investigations on similar Na concentrations have also reported such tendency of Na ions to occupy the Na2 sites; a Na1/Na2 ratio of 0.2 for a Na concentration of $85.71\%$\cite{Meng2008} and a Na1/Na2 ratio of 0.428 for a Na concentration of $83\%$.\cite{Alloul2008}

    For the \ch{Na_{0.75}CoO2} system, two Na ions were removed from each of Na layers to create the desired concentration.
    After structural optimization, we found that four Na ions occupy Na1 sites while eight other occupy Na2 sites.
    When compared with the previous DFT works, our Na1/Na2 is smaller than the previously reported ratio of one, therefore our calculations refers to a new ground state for Na concentration of $0.75\%$.
    Furthermore, for those Na at the Na2 sites the position of Na ions on the basal plane is slightly off the oxygen axes conforming to earlier experimental observations of H1 structure for this level of Na concentration.\cite{Huang2005}
    In the \ch{Na_{0.75}CoO2} system, in each of Na layers two Na ions moved from Na2 site to Na1 site to create a multi-vacancy cluster as marked by green enclosures in Fig. \ref{fig:2}(c).
    As a result, the \ch{Na_{0.75}CoO2} system had higher orthorhombic symmetry of the $PMMN$ space group with 16 symmetry operations.
    The occurrence of highly ordered Na vacancy clusters in \ch{Na_{0.75}CoO2} has previously observed in neutron diffraction experiments.\cite{Morris2007, Roger2007}
    Nonetheless, the stability of the ground state configuration was not as strong as the ground state configuration of the \ch{Na_{0.875}CoO2} system.
    For instance, a configuration with 6 Na ions at Na1 site had a slightly higher total energy of 2.6 meV while a configuration with no Na ion at Na1 site had a total energy higher by $59\si{.meV/Co}$.
    This implies that charge order transition for this system occurs at far lower temperature.

    The \ch{Na_{0.625}CoO2} system was created by removing 6 Na ions from the \ch{Na_{}CoO2} structure.
    After geometry optimization, we found that 6 Na ions were placed at Na2 sites while 4 other Na ions were placed at Na1 sites creating a structure that had the less symmetric orthorhombic $PMMA$ space group.
    The stability of the ground state in this structure is demonstrated by the fact than when one Na ion moves from Na1 site to Na2 site, the energy penalty is 15 meV while if a Na ion moves from Na2 site to Na1 site the energy penalty is 51 meV.
    Finally, in the \ch{Na_{0.05}CoO2} system, we found that four Na ions occupy Na2 site while for other occupy Na2 site.
    The system exhibits an orthorhombic symmetry with $PMMA$ space group.
    The stability of the ground state is demonstration by the fact that for a configuration with all Na ions on Na2 sites the energy penalty is 55 meV while there was no local minima for a configuration with all Na occupying Na2 sites.

    In order to examine the magnetic ordering of Co ions in the \ch{Na_{x}CoO2} system, we calculated the total energy of the ground state configurations for any given Na concentration ($x$) for three different spin alignments.
    First, the nonmagnetic state, second, the ferromagnetic (FM) state in which the magnetic moment of all Co ions were set parallel and finally the antiferromagnetic (AFM) state in which the magnetic moments of Co ions of any given \ch{CoO2} plane were set parallel among themselves while antiparallel to the adjunct \ch{CoO2} layer creating type A antiferromagnetic state.
    We chose the type A antiferromagnetism because it has been widely reported to occur in \ch{Na_{x}CoO2}\cite{Alloul2008, Bayrakci2005, Sushko2006, Bobroff2006} for various quantities of $x$.

    The nonmagnetic state was the most stable --having the lowest energy-- only in the case of full Na occupancy of the \ch{Na_{}CoO2} system.
    This was anticipated as in this system there is a \ch{Na+} for every Co ion forcing all Co ions in the system to have a valence state of 3+.
    Since \ch{Co^3+} is most stable in the low spin state where all lower $t_{2g}$ orbital are occupied by six electrons resulting in net zero spin.

    \begin{figure}
        \centering
        \includegraphics[width=0.95\columnwidth]{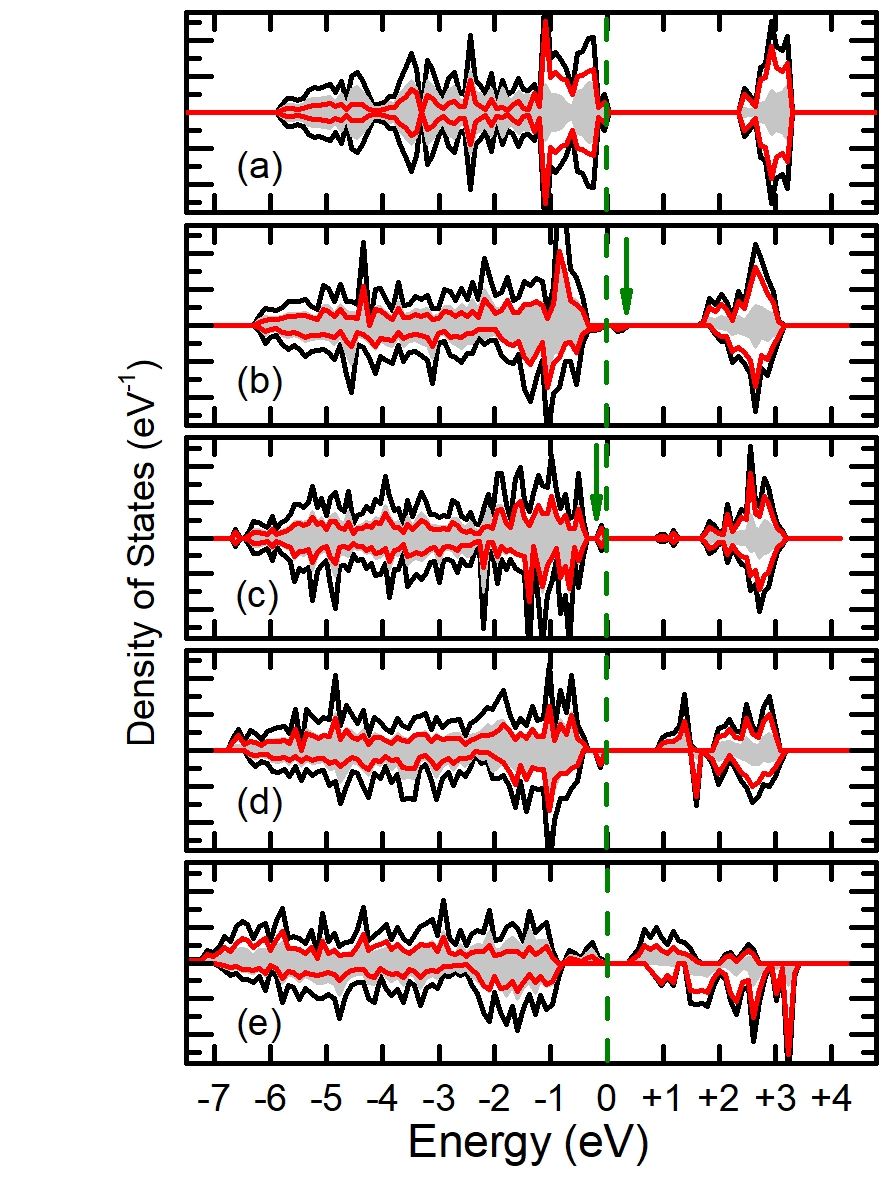}
        \caption{\label{fig:3}The total and partial density of states (P/DOS) of (a) \ch{Na_{}CoO2}, (b) \ch{Na_{0.875}CoO2} and (c) \ch{Na_{0.75}CoO2}, (d) \ch{Na_{0.625}CoO2} and (e) \ch{Na_{0.50}CoO2} are presented. The black and red lines represent the total and Co $3d$ states respectively. The gray shaded area represents O $2p$ states.}
    \end{figure}

    In the \ch{Na_{0.875}CoO2} system, the FM state was slightly more stable than the AFM state.
    Here we define $\Delta E = E_{\mbox{\scriptsize{AFM}}} – E_{\mbox{\scriptsize{FM}}}$ to quantify the magnetic phase stability.
    Larger $\Delta E$ values indicate a more stable ferromagnetic phase.
    In the case of \ch{Na_{0.875}CoO2}, $\Delta E$ was found to be $3.0\si{.meV/Co}$ which indicates that magnetically ordered phase may occur only at very low temperatures.
    We also found that the two net spin bearing Co ions in the system that were located in the proximity of the Na vacancies (Fig. S1).
    This can be understood if we consider the fact that by removing a \ch{Na^{+}} from the supercell, one Co ion changes its valence state from 3+ to 4+ to maintain charge neutrality.
    \ch{Co^4+} has only 5 electrons in the $t_{2g}$ orbitals thus leaving one electron unpaired.
    In the \ch{Na_{0.75}CoO2} system, the AFM state was considerably more stable with $\Delta E$ being $–19.2\si{.meV/Co}$.
    This is in agreement with previous experimental observation of the antiferromagnetic phase for $0.75 < x < 0.85$ with $T_N$ of $\sim 22 \si{.K}$.\cite{Bayrakci2005, Alloul2012}
    The origin of strong out of plane ($c$ direction) AFM order has been attributed to the strong second neighbor coupling of \ch{Co}--\ch{Na}--\ch{Co}.\cite{Johannes2005}
    For the \ch{Na_{0.625}CoO2} the FM phase was feebly stable with $\Delta E$ of $0.20\si{.meV/Co}$ while the \ch{Na_{0.50}CoO2} system had a $\Delta E$ of $4.0 \si{.meV/Co}$ indicating stronger FM alignment among Co ions which is in agreement the observation of FM behavior for $x = 0.65$ with $T_c = 68 \si{.K}$.\cite{Alloul2008}
    For Na concentrations intermediate to the ones studied here, a superposition of these structures might constitute the most stable Na patterning resulting in the phase separation phenomenon.\cite{Shu2009}

    To probe the electronic properties, the total and partial density of states (P/DOS) of \ch{Na_{x}CoO2} systems was calculated and is presented in Fig. \ref{fig:3}(a)-(e).
    As in Fig. \ref{fig:3}(a), the DOS of \ch{Na_{}CoO2} is fully symmetric with respect to the spin direction which indicates the nonmagnetic ground state of Co ions in this system.
    Furthermore, the fundamental bandgap is $\sim 2.2\si{.eV}$ which equals to the crystal field splitting between Co $t_{2g}$ and $e_g$ electrons in \ch{CoO2} system.\cite{Singh2000}
    The P/DOS of \ch{Na_{0.875}CoO2} system is presented in Fig. \ref{fig:3}(b) which also indicates a bandgap of $\sim 2.2\si{.eV}$ similar to the \ch{Na_{}CoO2} system.
    However, for \ch{Co^4+}, in addition to the crystal field splitting, there is a $p$-$d$ exchange splitting that pushes a portion of Co $t_{2g}$ (indicated by a green arrow) states into the fundamental band gap near the valence band maximum.
    The band insulating nature of \ch{Na_{x}CoO2} systems near Na saturation limits has been previously observed in experiments.\cite{Boehnke2014}
    The P/DOS of \ch{Na_{0.75}CoO2} system is presented in Fig. \ref{fig:3}(c).
    Here, due to the higher order crystal symmetry $t_{2g}$ electron of magnetic \ch{Co^4+} are further spilt into two subsets as indicated by green arrows in Fig. \ref{fig:3}(c).
    The split states are advanced in the fundamental band gap thus pushing the Fermi level closer to the conduction band minimum.
    For \ch{Na_{0.625}CoO2} and \ch{Na_{0.50}CoO2} systems as presented in Fig. \ref{fig:3}(d) and (e), the Fermi level touches the valence band resulting in metallic state which is in agreement with the previous experimental reports of metallic conduction for $x = \,\sim\!0.5$.
    As demonstrated in previous section the FM stability of \ch{Na_{0.50}CoO2} system ($\Delta E = 0.20\si{.meV/Co}$) is enhanced when compared with the one of \ch{Na_{0.625}CoO2} system ($4.0\si{.meV/Co}$).
    The ferromagnetic enhancement is positively correlated with number of Na vacancy in these systems, which indicates that ferromagnetism for lower Na concentration of the \ch{Na_{x}CoO2} system is carrier-mediated.

    In conclusion, by using pseudopotential and all-electronic density functional theory, we studied the electronic and crystal structure of \ch{Na_{x}CoO2} ($x = 1$, 0.875, 0.75, 0.6250 and 0.50).
    We found that in the \ch{Na_{}CoO2} and \ch{Na_{0.875}CoO2} systems, all Na ions occupy Na2 sites thus reducing the electrostatic repulsion between Na and Co ions.
    For lower concentration of \ch{Na_{0.75}CoO2}, \ch{Na_{0.625}CoO2} and \ch{Na_{0.5}CoO2}, we found that some of Na ions moved to the Na1 sites in order to decrease the in-plane \ch{Na}--{Na} electrostatic repulsion.
    Magnetically, there was strong tendency for type A antiferromagnetism in the \ch{Na_{0.75}CoO2} system with $\Delta E = –19.2\si{.meV/Co}$, while all other Na deficient systems had a weaker ferromagnetic tendency.

    \begin{acknowledgements}
        This work was supported by Japanese Society for Promotion of Science. The simulation facility was provided by Intersect Australia Limited.
    \end{acknowledgements}

    \bibliography{Paper3}{}
\end{document}